\documentclass[prb,aps,showpacs,preprint]{revtex4}

\usepackage[dvips]{graphicx}
\usepackage{bm}
\usepackage{pgf}
\usepackage{amssymb}

\begin{document}

\title{First-Principles conductance of nanoscale junctions from the polarizability of finite systems}
\author{Matthieu J. Verstraete$^{1,3}$}
\email{matthieu.jean.verstraete@gmail.com}
\author{P. Bokes$^{1,2,3}$}
\author{R. W. Godby$^{1,3}$}
\affiliation{
$^{1}$ Dept. of Physics, University of York, Heslington, York YO10 5DD, United Kingdom \\
$^{2}$ Department of Physics, Faculty of Electrical Engineering and Information
Technology, Slovak University of Technology, Ilkovi\v{c}ova 3, 812 19
Bratislava, Slovak Republic \\
$^{3}$ European Theoretical Spectroscopical Facility (ETSF)}

%%\pacs{63.20.-e,71.15.Mb,73.21.Hb,73.22.-f}

\begin{abstract}
A method for the calculation of the conductance of nanoscale electrical junctions is
extended to ab-initio electronic structure methods 
%% P.B. 12/11/2008
% ... but perhaps the abstract needs more work overall?.
which make use of the periodic supercell technique, 
and applied to realistic
models of metallic wires and break-junctions of sodium and gold. The method is
systematically controllable and convergeable, and can be straightforwardly
extended to include more complex processes and interactions. Important issues
about the order in which are taken both the thermodynamic and the static (small
field) limits are clarified, and characterized further through comparisons to model
systems.
\end{abstract}

\maketitle

\section{Introduction}
\label{intro_sect}

Nanoscale and molecular electronics is one of the most active topics of
research in physics today\cite{venkataraman_2006_conductance_exp_conformation,Quek07,
venema_2008_organic_electronics}; it has very important consequences, both for
fundamental research and for industrial processes, which will reach their
inherent quantum limits in the decade to come. Accurate methods of theoretical
as well as experimental characterization are essential, and many open questions
remain about the structure, equilibrium, and dynamics of nanometer-sized
systems carrying electrical currents. The ab initio%% \cite{kohn_1965_DFT_LDA}
simulation of materials properties is in a unique position to develop the scope
and our understanding of electronics at the nanoscale, providing the only
method of systematic analysis of electronic and structural effects, which are
never all simultaneously accessible in experiment. Present simulations of
nanoscale transport usually fall into two categories, either employing
time-dependent density functional theory
(TDDFT)\cite{kurth_2005_tddft_transport,Bushong05,Qian06} or various flavors of
Landauer-B\"uttiker-like formulas\cite{DiVentra00,Evers04,Koentopp08},
sometimes using a non-equilibrium Green's function
formalism\cite{keldysh_1965_negf} (for a review %of the basics of transport calculations 
see Ref.~\onlinecite{Koentopp08}). Due
to the way they are formulated, using embedding schemes, these techniques often
rely on localized functions (atomic\cite{Brandbyge02,Evers04,Toher05,Choi07,beste_2008_basis_set_effect_transport}
or Wannier\cite{ %wannier_1955_functions_1,wannier_1956_functions_2
Calzolari04,Strange08} functions) to
describe the single particle wavefunctions of the system. This introduces an
inherent difficulty in converging calculations, as the basis sets usually
cannot be refined systematically. There is still a great deal of
uncertainty about the precision of both experiments, e.g. due to fluctuation in
experimental conditions of contact and current flow, and theory, where no
standard model is yet accepted as being predictive of experiments, apart from
simple cases of continuous contact with conductances of at least one quantum of
conductance ($G_0$). The role and importance of electron-electron\cite{Delaney04,Ferretti05,Cehovin08,
Thygesen08, Myohanen08, Koentopp08} and electron-vibration\cite{Verdozzi06,Frederiksen07,Galperin08} 
interactions, has been recognized as an important factor in obtaining the correct transport 
properties even though satisfactory treatment for a general system at the ab initio 
level is not yet available. 

In the following, we present a method to calculate the conductance of a quantum
junction, which can be systematically converged and extended to include the
effects of different interactions. The formalism has been previously applied to
model systems in 1D\cite{bokes_2004_conductance} and jellium
slabs\cite{bokes_2007_four_point_conductance}, and is here extended to
incorporate 3D realistic ab initio electronic structures at the level 
of local or semi-local time-dependent density-functional theory. Section
\ref{method_sect} describes the method and how it must be adapted to suit the
periodic boundary conditions and supercells which are often used in ab initio
calculations. Preliminary numerical characterization is carried out in Section~\ref{num_char_sect}.
Section \ref{sodium_wire_sect} analyzes the convergence
properties of the method and applies it to monatomic wires of sodium. In
particular, the possibility of obtaining precise calculations of very low
conductances is explored for a tunnel junction. Finally, Section
\ref{gold_wire_sect} examines contact geometry and bonding effects for bulk
gold electrodes.

\section{Methodology}
\label{method_sect}

The conductance of a nanojunction characterizes the long-time dynamics 
of the electronic response of electrons to a driving electric field~\cite{bokes_2007_four_point_conductance}.
While formally the long-time limit demands the study of an extended system, for 
calculation purposes it is possible to consider a finite model for a finite time 
and the resulting conductance is obtained by extrapolation of the {\it conductance function}
\begin{equation}
	G^{2P} = \lim_{\omega \rightarrow 0^+} \lim_{L\rightarrow \infty} 
		G^{2P}(\omega,L), \label{eq-2-1}
\end{equation}
where $G_{2P}$ is the two-point conductance (see Ref.~\cite{bokes_2007_four_point_conductance}). 
The order of limits is important here: the one given characterizes transport 
in an extended system where as the reverse would reflect damped oscillations
of density in a finite (even though large) system. The underlying finite system 
can fulfill any boundary conditions and these do not affect the extrapolated results.
We take advantage of this fact and use the periodic boundary conditions 
and a plane-wave basis for the {\it ab initio} calculations below.

It is presumed in the following that the junction whose conductance we are
searching for is centered at 0 in the middle of the cell (which thus extends
from $-L/2$ to $+L/2$).

The calculation of the electronic response function at the level of local or semi-local TD DFT, 
which then leads to the conduction function (Eq. \ref{eq-2-1}), proceeds in three steps. First 
we perform a calculation of the occupied ($e_n < 0$) as well as 
the unoccupied ($e_n > 0$) eigenenergies $e_n$ and eigenstates $\phi_n(\bm r)$
of the system. 

Second, we use the eigenvalues and eigenstates to compose the positive imaginary-time Matsubara 
Green's function
\begin{equation} 
	\mathcal{G}(\bm r, \bm r';\tau) = \sum_{n} \phi_n(\bm r) \phi_n^*(\bm r')
	\frac{e^{-e_n \tau}}{e^{\beta e_n} + 1}  \label{eq-G-phiphi}
\end{equation}
As $\mathcal{G}$ is anti-periodic in imaginary time (fermionic) there is no need to
specify explicitly its behavior for negative imaginary time. 

The electronic response to the total electric field is characterized by
the polarizability 
\begin{eqnarray} 
	P(\bm r,\bm r';\tau) &=& - \mathcal{G}(\bm r, \bm r';\tau) 
		\mathcal{G}(\bm r', \bm r; -\tau) \nonumber \\
               &=&  \mathcal{G}(\bm r, \bm r';\tau) 
                \mathcal{G}(\bm r', \bm r; \beta -\tau) \label{eq-P-GG}
\end{eqnarray}
which, after Fourier transformation, $\tau \rightarrow \omega$, and 
integration over the cross-sectional area of the junction, $A$, gives the 
integrated polarizability relevant for charge transport
\begin{equation} 
	P(x,x';i\omega) = \frac{1}{A^2} \int \int dS_{\perp} dS'_{\perp}
	P(\bm r,\bm r';i\omega).
\end{equation}

%% P.B. 12/11/2008
% added the below paragraph for explanation of the statements 'can include further ...'
Methods going beyond the present level of approximations, i.e. using 
nonlocal exchange-correlation kernels\cite{Sai05,Jung07} or Green's-function-based
many-body methods\cite{Onida02}, would differ in the above Equations \ref{eq-G-phiphi} and
\ref{eq-P-GG}. The expression for the irreducible polarizability, Eq.~\ref{eq-P-GG}, 
would contain further vertex diagrams\cite{Onida02} and, in the case of many-body methods, 
the Green's function cannot be expressed in terms of one-electron wavefunctions as 
in Eq.~\ref{eq-G-phiphi}. However, the discussion that follows would apply also to these 
computationally more demanding approaches.

Finally, the third step consists of integrating the polarizability to 
obtain the conductance function.
For an infinitely long system, the 
conductance is obtained from the expression\cite{bokes_2007_four_point_conductance}
\begin{equation}
        G^{2P}(\omega,\infty) 
	= \omega \int_{-\infty}^{0} \int_0^{\infty} P(x,x';i\omega) dx dx', \label{eq-G-P}
\end{equation}
where this integral converges for any finite $\omega$ since $P(x,x';i\omega)
\rightarrow 0$ for $x,x' \rightarrow \infty$.  The integration region
corresponds to choosing elements of the polarizability which connect points on
opposite sides of the junction. This is intuitive, as we are interested in how a
perturbation on one side can influence charges on the other side, through the
junction.

\begin{figure}
\includegraphics[width=0.44\linewidth]{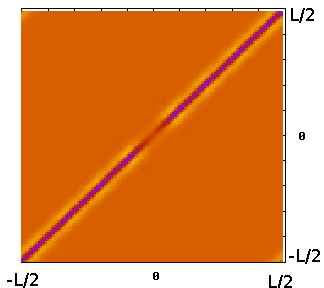}
\includegraphics[width=0.49\linewidth]{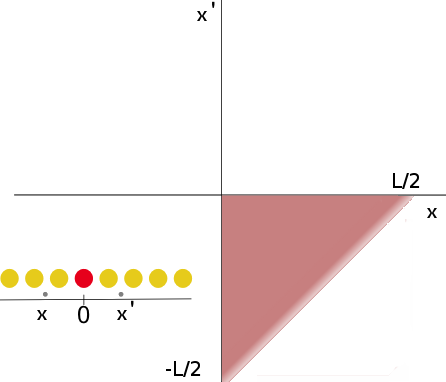}
\caption{\label{img_integration_region}Left: example of a color plot of the polarizability
$P(x,x')$ for a junction. $P$ is non zero only near the diagonal $x=x'$, and lower in the
central tunneling region (see below). Right: the region of spatial integration for
the polarizability as in Eq.~\ref{eq-G-omega-L}. For periodic systems the
polarizability will have spurious images, which must be excluded from the
integration region. Taking the thermodynamic limit will increase the size L of
the system, and the triangular domain converges uniformly to the quarter plane
$0<x<\infty$ and $-\infty<x'<0$. The inset is a cartoon of an atomic wire with a
central site and two positions, $x$ and $x'$.}
\end{figure}

For a finite system of length $L$ we obtain the corresponding function 
\begin{equation} 
	G^{2P}(\omega,L) = \omega \int_{\mathcal{D}} P(x,x';i\omega) dx dx',
	\label{eq-G-omega-L}
\end{equation}
where $\mathcal{D}$ is a domain of positive $x'$ and negative $x$ which must
guarantee the correct limiting procedure. Using
a periodic supercell, for $x'-x \rightarrow \pm L$, we approach a periodic
image of the system we want to study.
Further, if the system is not translationally invariant we will approach a region
of the $(x,x')$ plane (the lower right hand corner of Fig.~\ref{img_integration_region})
where the polarizability behaves very differently from
that near $(0,0)$ (typically one with a more metallic behavior and 
larger polarizability than the junction). Hence, a correct integration needs to
truncate the quadrant defined by $-L/2 < x < 0$ and $0 < x' < L/2$. For a
finite system there is no unique choice of $\mathcal{D}$, but there is a
natural one, which is $0 < x' < L/2$ and $-|x'| < x < 0$, or equivalently $0 <
x'-x < L/2$, defining a triangle between 0 and the points $(0,L/2)$ and
$(-L/2,0)$ (Fig.~\ref{img_integration_region}).

%In our work we have employed $\xi=1/2$; the behaviour for other values 
%is demonstrated in Fig~\ref{fig-1}b. \\
%\#\# Matthieu, do you have such a figure? \#\# \\
%
%  do we really need it? the math is pretty clear, the other domain is at best
%   the periodic copy or symmetric image
%

The finite size of the system determines the minimal frequency which can be
reliably described in the conductance, or equivalently the longest time
propagation. For longer times or lower frequencies the electrons will reach
the limits of the system, and the conductance decays.
The minimum frequency may be estimated as:
\begin{equation}
  \omega_{min} = 2 \pi v_F /L,
  \label{eq-minfreq}
\end{equation}
where $v_F$ is the Fermi speed. Thus, an electron at the Fermi level takes time
$1/\omega_{min}$ to traverse the whole system. This frequency will be essential
in determining how to extrapolate the conductance function to zero frequency.

We apply the formalism described above to systems described by modern
electronic structure methods. 
Many of these techniques use periodic boundary conditions to describe
crystalline structures, and represent wavefunctions and electronic densities
using plane wave basis sets. Here we describe the corresponding small changes
needed in the formalism.

First, as we wish to describe an isolated nanojunction between two leads (which
are in principle infinite) we will use only the zone-center $\Gamma$ k-point
of the Brillouin zone (BZ) along the axis of conduction. Using several k-points
would in effect simulate an array of interfering junctions. The
periodic boundary conditions are nevertheless exploited as the regions near the edge
of the simulation cell are described continuously instead of being brutally cut
off or terminated with hydrogen atoms. The thermodynamic limit along the
conduction axis must still be ensured by increasing the longitudinal system
size until the conductance converges.

In the directions perpendicular to $x$ a denser k-point grid can be used if bulk
three-dimensional leads are considered: the electronic structure of the leads
will thus be represented correctly, but care must be taken that the transverse
distance between images of the ``junction'' part of the cell is sufficient to
avoid interference between periodic images. In the case of a purely 1D system
no perpendicular k-points are necessary as the system is supposed to be isolated
in vacuum along $y$ and $z$.

\section{Numerical characterization}
\label{num_char_sect}
\begin{figure}
\includegraphics[width=0.9\linewidth]{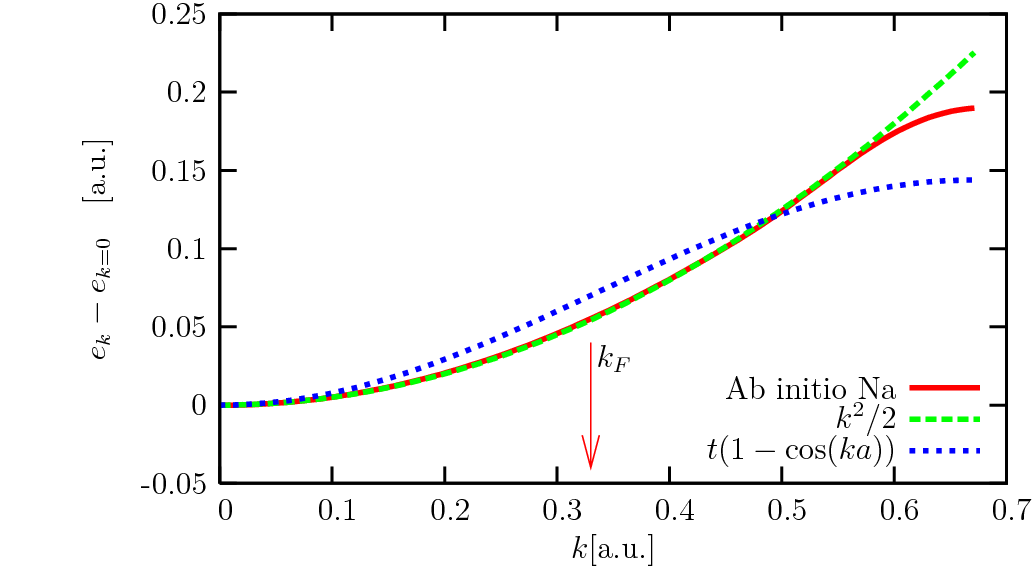}
\caption{\label{dispersions} The dispersion of eigenenergies, obtained from 
the {\it ab initio} calculations (see Sec.~\ref{sodium_wire_sect} for details), is almost 
identical to that of 1D jellium using the bare electron mass. 
The TB model is fit to have the same Fermi speed, and the differences with
respect to the {\it ab initio} dispersion are more significant.}
\end{figure}

In order to understand the convergence behavior of the main results we have also analyzed
a finite 1D jellium model and a finite 1D tight-binding model. The length $L$ of both
systems can be made much larger than in the {\it ab initio} models,
which allows detailed study of the extrapolation to small frequencies.
Similarly to the {\it ab initio} case, both models use periodic boundary 
conditions and their parameters are such that the density 
and the Fermi speed of the particles will be identical to that of a sodium chain 
studied within the self-consistent {\it ab initio} calculations. In fact, the jellium 
model with a bare electron mass is an excellent model for the sodium 
wires. This is shown in the Fig.~\ref{dispersions} where the 
dispersion of eigenenergies, obtained from the {\it ab initio} calculations,
is compared with the dispersion of the 1D jellium and the fitted TB model. 
More generally, these two models represent two extremal types of electronic structure 
for metallic wires. The correct understanding of the extrapolations to infinite 
size and zero frequency  of their conductance functions is very useful for performing 
extrapolations of more realistic but numerically more demanding {\it ab initio} calculations.

For both model systems one can find the eigenstates exactly by going into 
reciprocal space. The conductance function~(Eq.~\ref{eq-G-omega-L}) 
%% takes the form
%% p.b. perhaps I could put the derivation of the conductance function 
%% into the appendix?
%In contrast to Eq.~\ref{eq-G-P}, the conductance function is obtained by summing over 
%the discrete lattice sites
%\begin{equation} 
%	G^{2P}(\omega,N) = \omega \sum_{- \xi N/2}^0 \sum_{0}^{\xi N/2-1} 
%	P_{nm}(i\omega),
%\end{equation}
%%and
can be expressed using the exact eigenstates of the Hamiltonian
\begin{eqnarray} 
        G^{2P}(\omega,L) = 2 \omega \sum_{ij} |s(p_i-q_j)|^2
                        (1-n_{q_j})n_{p_i} \nonumber \\ 
			\times \frac{e^{\beta(e_{p_i} - e_{q_j})}-1}{
                        i\omega + e_{q_j} - e_{p_i}}, \label{eq-G-N}
\end{eqnarray}
where the sum goes over all eigenstates, $p_i$ or $q_i$ are the momenta of the 
eigenstates, $n_{p_i}$ are Fermi occupancies of the state $p_i$ at temperature $T=1/k_B \beta$,
$e_{p_i}$ are the eigenenergies corresponding to an eigenstate with 
the momentum $p_i$, and the factor 2 accounts for the spin degeneracy. The function $s(~)$
represents the conductance vertex-factor (similarly to the expression for the conductivity 
in terms of the polarization function, see e.g. Bruus and Flensberg\cite{bruus_2004_many_body_book})
and takes different forms for the two systems, as given below.

The jellium model consists of a 1D non-interacting electron gas of total length 
$L$ and density $n=N/L$ where $N$ is the total number of electrons. 
The eigenvalues are 
\begin{equation} 
	e_{p_i} = \frac{p_{i}^2}{2} - E_F, \quad p_i=\frac{2\pi}{L} i, i=\pm 1, \pm 2, \ldots 
\end{equation}
and the conductance vertex-factor is 
\begin{equation} 
	s(p) = \frac{1}{\sqrt{2}L} \int_0^{L/2} dx e^{-ipx} = 
	    - \frac{i}{\sqrt{2} L p} \left( 1 - e^{-ipL/2} \right).
\end{equation}
The Fermi energy is obtained from the requirement that the charge per unit length is identical 
to that of the sodium wires; the length $L$ is set to $N d_{Na}$, where 
$d_{Na}$ is the inter-atomic distance of the sodium wire so that 
the density of electrons is $n=N/L=1/d_{Na}$.

In the case of the tight-binding model, the Hamiltonian has the form
\begin{equation}
        H^{TB} = \sum_{-N/2}^{N/2-1} -\frac{t}{2} \label{eq-H-lattice}
                \left( c^{\dagger}_n c_{n-1} + h.c. \right),
\end{equation}
with the resulting eigenvalues $e_{p_i} = - t\cos(p_i)$ for a state with 
momentum $p_i$. The momentum only takes discrete values
\begin{equation} 
	p_i = \frac{2\pi}{N} i, \quad i=0,1,2,...,N-1
 \end{equation}
and the conductance vertex-factor is
\begin{equation} 
	s(p) = \frac{1}{\sqrt{2}N} \sum_{n=0}^{N/2} e^{ip n} 
		= \frac{1}{\sqrt{2}} \frac{e^{ip(N/2+1)} -1 }{e^{ip} -1}.
\end{equation} 
At half filling, the Fermi momentum $k_F=p_{(N/4)}$
and the Fermi speed is $v_F=de(p)/dp=t$. Making use of the Fermi speed of the sodium wires 
considered in Sec.~\ref{sodium_wire_sect} $v_F=0.33$ a.u., we identify the TB parameter 
as $t=v_F/d_{Na} \approx 0.07$.

\begin{figure}
\includegraphics[width=0.9\linewidth]{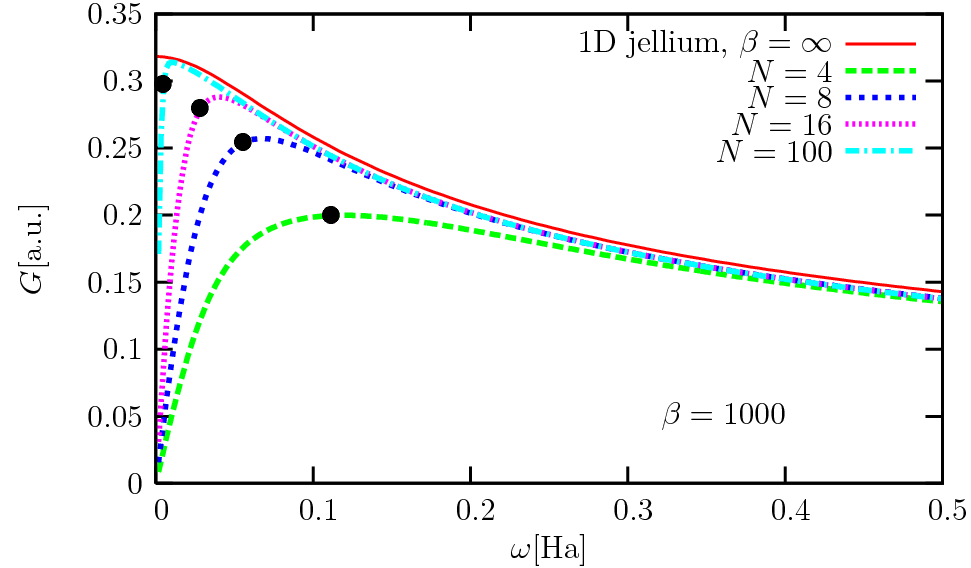} \\
\includegraphics[width=0.9\linewidth]{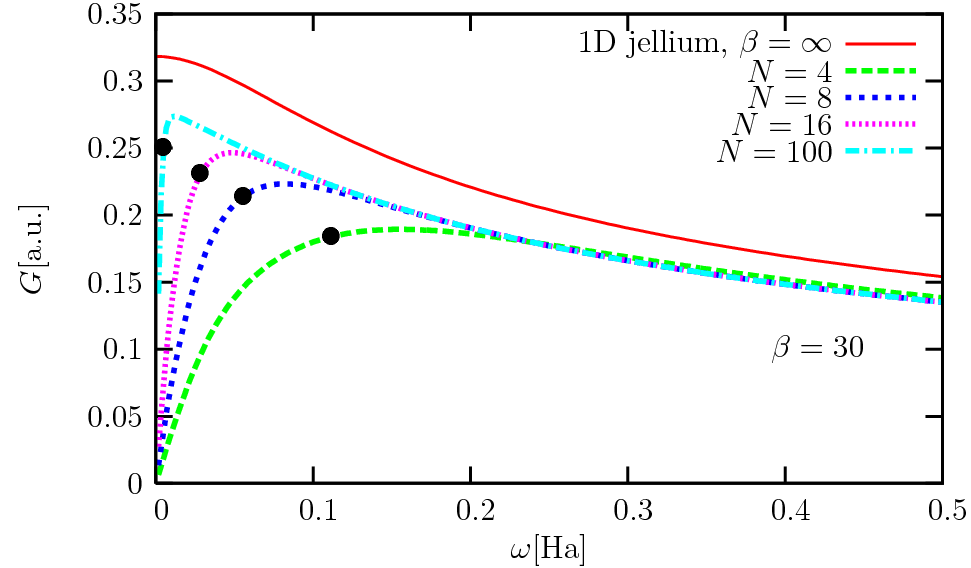}
\caption{\label{jellium_cond}Convergence of the conductance of a jellium wire
with respect to system size. The lengths correspond to sodium wires 4, 8, 16
and 100 atoms long. Upper graph: $T \ll E_F$ ($\beta=1000$). Lower graph: $T
\sim E_F$ ($\beta=30$).  The curves at low temperature approach the analytical
infinite-size result (for T=0 - continuous line). The curves should be extrapolated
to zero frequency, disregarding values for $\omega \lesssim \omega_{min}$ (indicated by a
dot on each curve).}
\end{figure}

The evaluation of the expression (\ref{eq-G-N}) for both models is very fast and
can be done for much longer wires than in the case of first-principles calculations.
In the top panel of Fig.~\ref{jellium_cond}, we show the conductance functions for jellium wires of lengths $L=N d_{Na}$
with $N=4,8,16$ and $100$ at temperatures much lower than the Fermi energy. The curves 
converge smoothly to the zero-temperature infinite-length limit that is known 
analytically~\cite{bokes_2004_conductance}, and readily give the static limit of one quantum of conductance
$G^{2P}(0,0) = 2e^2/h = 1/\pi$ a.u. Furthermore, the functional form shows finite size effects 
precisely according to the expected criterion (Eq.~\ref{eq-minfreq})
\begin{equation} 
	\omega_{min} = \frac{0.44}{N} = 0.11, 0.05, 0.03, 0.004
\end{equation}
for $N=4,8,16,100$ respectively. The extrapolation for $N=8$ or $N=16$ gives
good estimates of the conductance, through the limit in Eq.~(\ref{eq-G-omega-L}).
At temperatures comparable with the Fermi energy (bottom graph in Fig.~\ref{jellium_cond}) 
the extrapolated value is somewhat below the zero temperature limit but the functional 
form of the conductance is essentially identical.

\begin{figure}
\includegraphics[width=0.9\linewidth]{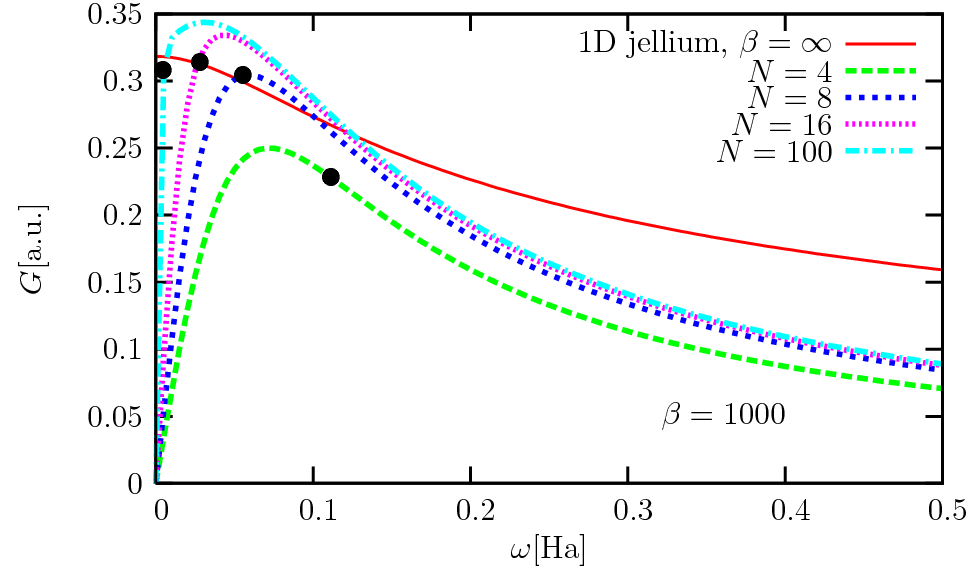} \\
\includegraphics[width=0.9\linewidth]{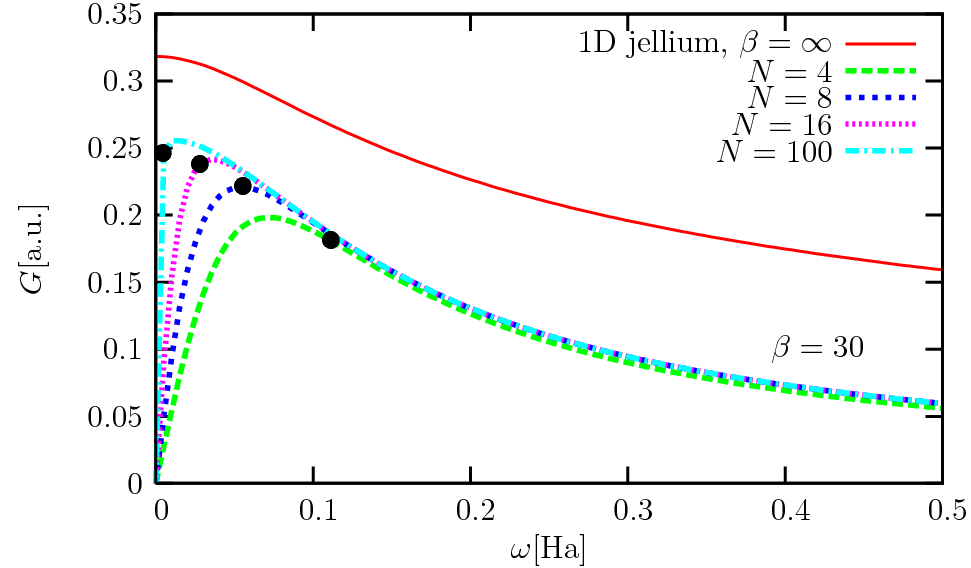} 
\caption{\label{tb_cond}Convergence of the conductance function of a
tight-binding model with respect to system size (4, 8, 16 and 100 atoms). $v_F$
is fixed to that of the sodium wire. Upper graph: $T \ll 2 t$ ($\beta=1000$)
Lower graph: $T \sim 2 t$ ($\beta=30$) where $2t$ is the bandwidth. The curves
give the correct zero temperature conductance for $L \rightarrow \infty$, but
the extrapolation to $\omega = 0$ is complicated by the non-monotonic behavior
at small frequencies.}
\end{figure}

On the other hand, the tight-binding model, shown in the Fig.~\ref{tb_cond},
offers less reliable extrapolations. This is caused by the non-monotonic behavior 
of the conductance function at small frequencies, which in turn arises because of the
bandwidth of the model (here the bandwidth is $2t=0.14$a.u.).
This non-monotonic behavior weakens if we look 
at the conductance curve at higher electronic temperature. This is shown in the 
lower graph in the Fig~(\ref{tb_cond}) for a temperature comparable to the bandwidth.
For these temperatures the conductance function can be easily extrapolated to (high-T)
conductance values identical to the jellium model (Fig.~\ref{jellium_cond}, lower graph).
This suggests that, in principle, by performing the calculations at different electronic
temperatures or smearing, one may enhance the extrapolation procedure in realistic 
calculations, which may combine aspects of free-electron (jellium) and localized electron
(TB) behaviors. 

To summarize, the models show good convergence in the extrapolated conductance
at $\omega=0$ for systems of length equivalent to 8 or 16 atoms. The
jellium dispersion is quite close to the {\it ab initio} one for Na, whereas
the tight binding one is not - this could be expected from the simple metal nature
of sodium. The conductance function of the TB chain is qualitatively different,
and actually overshoots the quantum of conductance for small imaginary
frequencies and large $L$. Finally, for high temperature (T comparable to
$E_F$) both models depart from the analytic curves for T=0 and the conductance
decreases.

\section{Sodium monowires: size convergence and energy dependencies}
\label{sodium_wire_sect}
We begin the \textit{ab initio} studies with a prototypical application: the calculation of the conductance of
a uniform monatomic wire (monowire) of sodium atoms. With one s electron per atom,
and given the simple-metal nature of sodium, the conductance in the
independent particle case will be two quanta of conductance, due to
spin-degeneracy.

\begin{figure}[htb]
\includegraphics[width=0.9\linewidth]{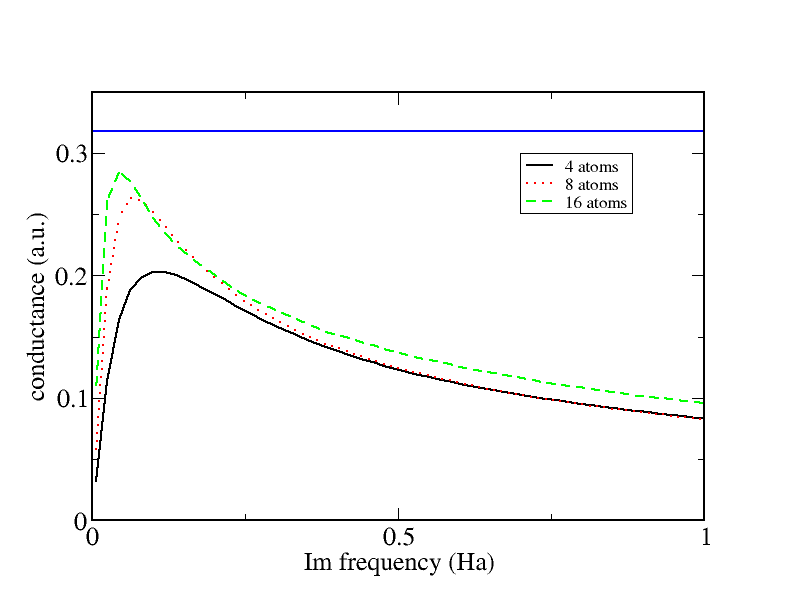}
\caption{\label{img_na_natom_conv}Convergence of the conductance of a
continuous monowire of sodium atoms, with respect to cell size (4, 8, and 16 atoms in the unit cell).
The minimum frequency for which the conductance function is valid depends on
the Fermi speed and goes down with increasing system size. For 8 atoms the
extrapolation to zero frequency is already good, arriving close to the expected 
2 quanta of conductance (horizontal line) for a non-interacting system with
spin-degeneracy.} 
\end{figure}
\emph{Technical details}\\
The ground state wavefunctions and electronic structure are calculated within
density functional theory (DFT)\cite{hohenberg_1964_DFT,kohn_1965_DFT_LDA},
using a plane wave representation, with the ABINIT\cite{ABINIT_short} or
SFHINGX\cite{boeck_sfhingx} codes (the results have been checked to be
independent of the ground state code used). We employ norm-conserving
Troullier-Martins\cite{troullier_1991_nc_psp} type pseudopotentials, with
nonlinear core corrections\cite{louie_1982_NLCC} and the d channel as a local
potential. The kinetic energy cutoff (20 Hartree) and number of bands (100 per
Na atom) were over-converged to allow for full checks of the convergence of the
conductance calculation. The calculation of the conductance, in a module of the
GWST code\cite{rojas_1995_GWST_prl}, was carried out with a kinetic energy
cutoff of 8 Ha.

The inter-atomic distance was set to 2.477~\AA. Other distances were checked, but
do not influence the results appreciably: the Fermi point for the wire is fixed
by the parabolic nature of the bands, and, most importantly in our case, the
Fermi speed scales as the inter-atomic spacing, making the critical minimum
frequency (Eq.~\ref{eq-minfreq}) independent of the spacing.

The perpendicular size of the unit cell is more important, as we wish to
simulate a truly 1D system using a supercell. The lateral dimensions of the
unit cell are fixed to 4~\AA, which is enough to ensure that conductance only
happens along the wire direction. Checks with 8 and 16~\AA{} cells showed that
the conductance function is already well reproduced to within a few percent.

\emph{Results for uniform Na monowires}\\
% figures
%  conductance as a function of length
Figure \ref{img_na_natom_conv} shows the size convergence of the conductance
function, for unit cells containing 4, 8, and 16 atoms. As can be seen in this
simple case the conductance function is correct to lower and lower frequencies
as the system size is increased, and the extrapolation tends towards $1/\pi$.
Already with an 8 atom cell the
linear extrapolation of $G^{2P}$ to zero frequency gives a value very close
to $1/\pi$, as expected from the previous section.
Comparing the values obtained for successively larger system sizes
gives an estimate of the residual error in converging $L$. A 16 atom unit cell
is about 40~\AA{} long. The very slow convergence of $G(\omega_{min}, L)$
as a function of $L$ is due to the 1D
character of the system. Coulomb screening in 1 dimension is quite inefficient
and the polarizability decays quite slowly. In 3D systems the screening will
be stronger and the size convergence quicker

A very important point is to have a good estimate of the Fermi level. In 1D
atomic chains this is not trivial, as the equivalent of k-point sampling is the
length of the system. The distance between the levels bracketing the $E_F$ decreases
with $L$, but not uniformly. For small $L$, there is an alternation in the position
of the Fermi level: for wire lengths which are multiples of 4 the Fermi level
is exactly on a single particle state, whereas for other values it is higher,
and between states. The dielectric response of these two cases is very
different, as one case appears to be a metal and the other an insulator (whose
gap goes to 0 as $L \rightarrow \infty$). In the interest of brevity we have
not included the results for $L= N d_{Na}$ with $N=5, 6, 7$ in
Fig.~\ref{img_na_natom_conv}: they oscillate slightly (as a function of $N$)
and converge more slowly, though to the same end result. We will see these effects
again in the next Section for the case of a wire with a gap.

In our formulation the conductance is expressed in imaginary frequency. This
implies contributions from all electron hole pairs in the polarizability, not just
those for states near the $E_F$. Because of this the convergence in the number of
states is comparable to (but slightly faster than) that of a GW calculation,
with between 5 and 20 bands per electron (for Na wires we need 10 bands per electron).

%%TEST THIS EXPLICITLY WITH X CONDUCTANCE OF BULK SYSTEM??? With 4 A between
%%wires not very different from bulk... no? Add perp k-points to see? Is
%%difference btw longitudinal interatomic distance (short) and perp (long)
%%important enough to give 1D behavior? Or is it just the k-points combined with
%%the cell dimensioning?

%  gap wires
%
\begin{figure}
\includegraphics[width=0.9\linewidth]{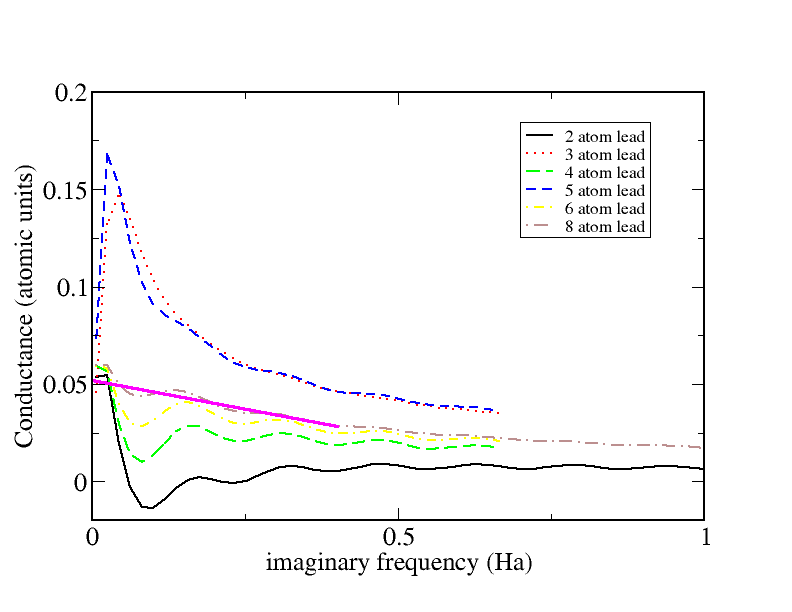}
\caption{\label{img_gap_na_natom_conv}Convergence of the conductance of a wire
with a gap (1 atom missing), with respect to the size of the lead wires (2 to 8
atoms in each lead). The conductance functions converge well for even numbers
of atoms in the leads, due to a correct positioning of the Fermi level, and can
be extrapolated to zero frequency (the line segment extrapolates from 8-atom lead case).
With odd numbers of atoms the Fermi level is positioned in an artificial gap.}
\end{figure}
\emph{Na monowires with a gap}\\
We now proceed with an inhomogeneous case: a wire of Na atoms with a gap (of
width d). This is the simplest example of a nanojunction. The conductance will
naturally go into a tunneling regime as the gap becomes wider. This example is
important firstly because it has a simple dielectric response,
but also because tunneling is an important and extreme
regime for the conductance. Fig.~\ref{img_gap_na_natom_conv} shows the size
convergence of a wire with a gap of one atom. The two remaining parts of the
wire are of equal length, increasing from 2 to 8 atoms (each).
The even length leads converge relatively quickly to a regular conductance
function, from below. The odd length leads converge from above but very slowly,
because the HOMO/LUMO states are quite far apart, which gives a badly placed
Fermi level, as above for continuous wires (bad in the sense that it is far
from the limiting Fermi level for an infinite system). Extrapolation to 0
frequency gives a conductance of 0.05 a.u. ($\pm 0.005$).

% fractional gaps
\begin{figure}
\includegraphics[width=0.9\linewidth]{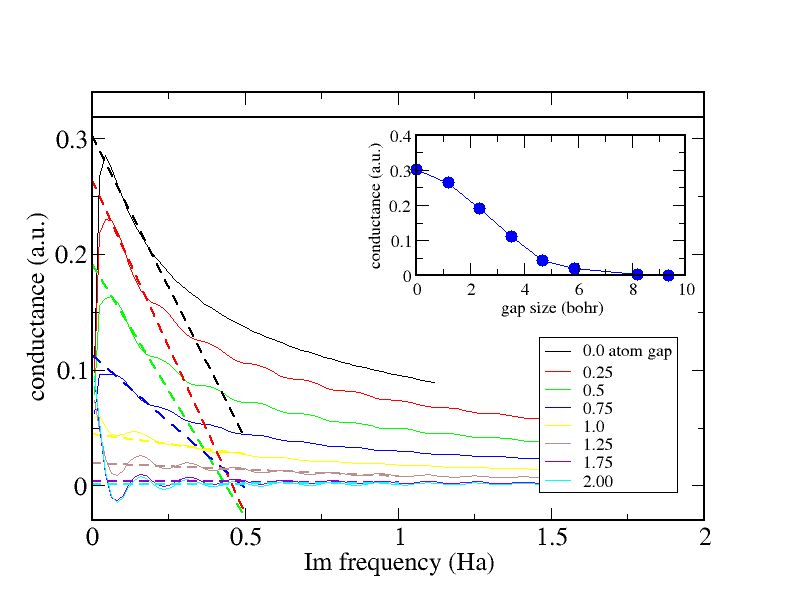}
\caption{\label{img_fract_gap_na}Conductance function of gapped wires, with
respect to imaginary frequency, and for different gap sizes (in fractions of
the inter-atomic distance in the regular wire). There are 8 atoms in each lead.
Inset: extrapolated 0 frequency conductance, as a function of gap length. The
conductance saturates for small gaps, and for gaps larger than $\sim$ 0.75
inter-atomic distances the tunneling decay appears.}
\end{figure}
The oscillations of $G(\omega)$ are due to aliasing effects in the Fourier
transform from imaginary time to imaginary frequency. These are in some cases
difficult to eliminate for very low amplitude elements of $P(x,x',i\tau)$.

Finally we consider the transition from the conducting to the tunneling regime,
by increasing the gap size (for 8 atom leads). The conductance functions are
represented in Fig.~\ref{img_fract_gap_na}, with the inset showing the decrease
of the extrapolated static conductance as a function of the size of the gap.
The conductance decreases with gap size, and by a gap of $3/4$ of an atomic
spacing (about 4 Bohr) the tunneling regime begins with exponential decay of
$G$.  The tail of $G(d)$ gives a very good fit to $G(d) = 1.68 \; exp(- 0.770
\; d)$ where $d$ and $G$ are in atomic units.
%%$G(d) = 1.68214 \; exp(- 0.769774 \; d)$ $d$ and $G$ are in atomic units.
The dashed lines are linear extrapolations of $G$ fit to the interval $[0.05,
0.2]$ Hartree. For the lowest curves (largest gaps) the fit was performed
further out, on $[0.2, 1]$, as the functions are flatter and the aliasing noise
more important. In this way we are able to represent quite small conductance
values, down to 0.001 a.u. (or 0.003 $G_0$).

A similar system was examined by Beste {\it et al.} in
Reference~\onlinecite{beste_2008_basis_set_effect_transport}, but with gold chains
instead of Na. They find a conductance of about 0.12 a.u. for a gap of
$d_{Au}/2$ (1.28\AA), which is close to our value of 0.18 a.u. for $d_{Na}/2$. One important
conclusion of Ref.~\cite{beste_2008_basis_set_effect_transport} is the very
strong deviations which can appear depending on the basis set nature, with
localized basis sets. Our results can be converged systematically, using
a plane-wave basis set, but are probably heavier calculations as a trade-off.

%It is however clear that for the
%study of certain experimentally interesting systems, whose conductance can go
%down to $10^{-5} G_0$ or even less(NEEDAREF), we will need a more robust
%algorithm and fit for the Fourier transform - this work is ongoing.

%  convergence as a function of bands? T?

\section{Gold wires: lead structure and k-point sampling}
\label{gold_wire_sect}
\begin{figure}
\includegraphics[width=0.9\linewidth]{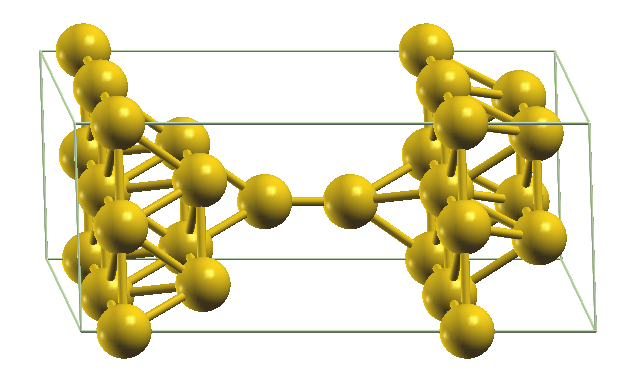}
\caption{\label{img_Au_unit_cell}Unit cell of a short gold wire contacted
with bulk gold electrodes, which are 2 layers thick. Both atoms sit at
natural FCC hollow positions on the (111) surfaces.}
\end{figure}

We now proceed with a more structured system, showing an explicit constriction.
A gold junction is made from a 2 atom wire contacted to bulk 3D electrodes. The
electrodes are FCC stacked gold (at the experimental nearest neighbor distance of
2.9~\AA{} of Ref.~\onlinecite{maeland_1964_au_acell}) of which we use a 2$\times$2
(111) surface unit cell. The wire atoms are evenly spaced with the FCC
inter-layer distance of 2.37~\AA{} (which is compressed compared to the DFT-LDA
equilibrium distance of 2.5~\AA{} for the infinite straight
wire\cite{sanchez-portal_1999_stiff_monatomic}). No relaxation of atomic
positions with respect to the bulk is taken into account, but the addition of
further structural effects is in no way more difficult; contrary to some other
approaches to transport\cite{beste_2008_basis_set_effect_transport,
kurth_2005_tddft_transport} we are not constrained to specific unit cell
lengths or layer spacings. A typical unit cell is shown in
Fig.~\ref{img_Au_unit_cell}, for the minimal electrode thickness of 2 layers
(in each electrode). Electrodes 3 and 4 layers thick were also tested. Because
of periodicity, and in order to maintain a continuous FCC structure at the cell
boundary, the point of contact of the wire to the right electrode alternates
between the different possible FCC stacking sites (for 2 and 4 layer
electrodes), and an on-top position (for the 3 layer electrodes).  We find
little effect of contact position on the conductance of the junction (see
below) in this continuously metallic, well contacted case.

\emph{Technical details}\\
% gold pseudopotential cite Andrea and co. 6s electron only.
The pseudopotential we use is of the Hartwigsen-Goedecker-Hutter
flavor\cite{hartwigsen_1998_psp_hgh}, with only the 6s electrons in the
valence. Our choice of pseudopotential is justified by its softness,
by the chemical homogeneity of the system, and our intention to go
beyond LDA and include many-body corrections. We have performed tests on FCC
gold in the GW approximation (which is beyond the scope of the present
article), including the 5d electrons. The problems recognized by Marini {\it et
al.}\cite{marini_2001_Cu_GW} for Cu appear for Au as well: the exchange
self-energy is quite badly represented for the 5d electron states, due to the
absence of the 5s and 5p. The latter are far in energy but have an important
spatial overlap with the 5d. Consequently, the exchange self-energy lacks
important contributions if one uses only the valence electrons. The GW d bands are very
poor (whereas their position in LDA is very close to experimental values);
some bands are pushed down and others up to the Fermi level, which would change
the conductance severely. The use of
a purely 6s potential is less realistic but reduces these exchange effects
(which are now between the 6s and the core 5d states). A more complete solution is
that adopted by Shishkin and Kresse\cite{shishkin_2006_GW_PAW} in the PAW
formalism\cite{blochl_1994_PAW}. As PAW allows explicit reconstruction of the
core states, the exchange with the valence can be calculated explicitly.
Finally, the d electrons do not complexify the independent particle conductance
calculation formally, but do make the calculations much heavier (with an
additional 10 electrons per atom). As the states at the Fermi level are purely
s-electron like, the conductance will not be affected strongly. However, as our
method is an integral of the dielectric response of the system, the
absence of the d electrons {\it will} have an indirect effect, through
changes in the polarizability.

\emph{Results}\\
From a calculation of a uniform wire (with k-points along the wire axis), we
estimate the Fermi speed in the wire to be 0.42 au ($1.9 \; 10^{6}$ m/s),
corresponding to a wavelength of 7.22 Bohr. A simple metal approximation for
the bulk gives an estimated Fermi speed of 0.64 au ($1.4 \; 10^6$ m/s) from the
Au Seitz radius. A DFT calculation of FCC bulk naturally gives a more complex
band structure - the modulus of the Fermi speed varies by some 20\% in
reciprocal space. The HGH pseudopotential gives an average value of 1.02 au
($2.2 \; 10^{6}$ m/s). A more complete pseudopotential with d electrons reduces
this value to 0.67 au ($1.5 \; 10^{6}$ m/s). The value we are interested in is
the speed of propagation of an electronic signal through the whole system, i.e.
through the 3D bulk (with the pseudopotential we are using) and the wire, which
will be between the pure bulk and pure wire values.
%
%  note: with hgh psp Au erroneously has a FS point along Gamma-L with SO and d
%  electrons there is no such point, it is pushed to the L-W branch
%
With the bulk and wire $v_F$, we can estimate the minimal frequencies which can be represented
for different unit cell sizes. With cells of lengths 27, 36, and 45 Bohr,
we obtain $\omega_{min} = $ 0.163, 0.132, and 0.113 Hartree.

%  conductance for thickening leads
\begin{figure}
\includegraphics[width=0.9\linewidth]{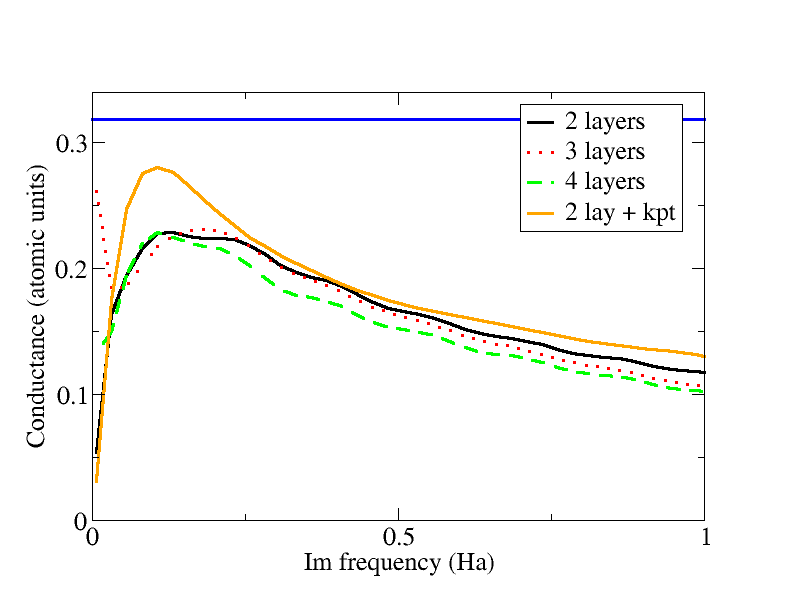}
\caption{\label{img_Au_thick_lead_conductance}Conductance functions of gold
junctions containing a 2 atom wire and bulk FCC leads. For 2, 3 and 4 layers of
gold in the leads, and only the $\Gamma$ k-point (solid black, dotted, and
dashed), and for 2 layers of gold and a 4$\times$4 sampling of the BZ
perpendicular to the wire axis (solid orange curve). The horizontal line
is the quantum of conductance. The peak in the conductance moves to lower
energies as the system length is increased. In the case with denser BZ
sampling the conductance function is much better represented, even if the
minimum frequency is not changed.}
\end{figure}
In Fig.~\ref{img_Au_thick_lead_conductance} (lowest three curves) we represent the
conductance as a function of imaginary frequency for a series of gold junctions
(like that schematized in Fig.~\ref{img_Au_unit_cell}) with 2, 3 and 4 layers of FCC gold
in the bulk leads. As before for linear wires, initially only the $\Gamma$ point
wavefunctions are used, and $G(\omega)$ takes similar values to the case
without leads, which would suggest a similar system length of ca. 8 layers
in each lead to converge the conductance. The extrapolated value, close to the 
quantum of conductance is in agreement with the results of the more extended models 
of the monatomic gold contacts~\cite{Jelinek08}.

With bulk leads it is essential to look at k-point convergence: to represent the
bulk states correctly, we increase the sampling of k-points in the direction
perpendicular to the junction axis. Again, as noted above, one must also take care
to keep the nanojunctions themselves well isolated in the perpendicular
direction, to avoid interference effects which would be amplified by the
perpendicular k-points. With 2 layer leads and a 4$\times$4 sampling of
k-points, we find the conductance function shown by the top curve of
Fig.~\ref{img_Au_thick_lead_conductance}. The low frequency behavior is now much
closer to that of the longer Na wires. Increasing the lead size thus has two
effects, which are controllable separately in this 3D case: first improving the
representation of the density of states (DOS) (which can also be achieved by
using the perpendicular k-points if the leads are bulk-like), and second
lowering the minimum representable frequency $\omega_{min}$ (which can only be
achieved by increasing the system length L). In the future, a more extended
study will combine the effects of longer leads and perpendicular k-point
sampling, but the computational load will require the parallelization of our
code, which is underway.

To summarize, the bulk leads on gold wires show that even with relatively small
system sizes a constriction limits the conductance to a single quantum of
conductance. Bulk 3D leads give much stronger screening than 1D ones, and faster
convergence of the conductance function. Including k-points to sample the
perpendicular electronic states in the leads improves the description of the
DOS and the screening.

%%Q: do a 2x2 kpt grid calculation? Would be nice for a semiconvergence study.

\section{Conclusions}
\label{conclusions_sect}
We present a new method to calculate the transport properties of nanoscopic
junctions. The extension of the basic formalism to periodic systems and
realistic electronic structure is detailed, and the convergence properties are
compared to model systems. Order-of-limits problems are reviewed, which also
concern many other approaches to quantum transport, as well as numerical
issues. The differences and inherent advantages of the method are discussed, in
particular the way it treats leads and its systematic convergeability in number
of single particle states and the spatial representation of different quantities.
Applications to sodium and gold nanojunctions is presented. The first show the
properties of purely 1D systems, and demonstrate the variation between regimes
of continuous metallicity and of tunnel junctions. The gold junctions
explore contact geometries and some of the fundamental differences between 1D
and 3D electrode structure and screening. 

\acknowledgements
The authors wish to acknowledge fruitful discussions with A. Ferretti, P.
Rinke, and C. Freysoldt. This research has been supported by the NANOQUANTA EU
Network of Excellence (NMP4-CT-2004-500198), the NATO Security Through Science Programme (EAP.RIG.981521) 
and by Marie Curie fellowship MEIF-CT-2005-024152. Some computer time was provided by the White Rose Grid.

\bibliography{JABREF_totalbib}

\begin{thebibliography}{10}

\bibitem{venkataraman_2006_conductance_exp_conformation}
L. Venkataraman, J. Klare, C. Nuckolls, H. M.S., and M. Steigerwald, Nature
  {\bf 442},  904  (2006).

\bibitem{Quek07}
S.~Y. Quek, L. Venkataraman, H.~J. Choi, S.~G. Louie, M.~S. Hybertsen, and
  J.~B. Neaton, Nano Letters {\bf 7},  3477  (2007).

\bibitem{venema_2008_organic_electronics}
L. Venema, Nature {\bf 453},  996  (2008).

\bibitem{kurth_2005_tddft_transport}
S. Kurth, G. Stefanucci, C.-O. Almbladh, A. Rubio, and E. Gross, Phys. Rev. B
  {\bf 72},  035308  (2005).

\bibitem{Bushong05}
N. Bushong, N. Sai, and M.~D. Ventra, Nano Lett. {\bf 5},  2569  (2005).

\bibitem{Qian06}
X. Qian, J. Li, X. Lin, and S. Yip, Phys. Rev. B {\bf 73},  035408  (2006).

\bibitem{DiVentra00}
M.~D. Ventra, S.~T. Pantelides, and N.~D. Lang, Phys. Rev. Lett. {\bf 84},  979
   (2000).

\bibitem{Evers04}
F. Evers, F. Weigend, and M. Koentopp, Phys. Rev. B {\bf 69},  235411  (2004).

\bibitem{Koentopp08}
M. Koentopp, C. Chang, K. Burke, and R. Car, J. Phys.: Condens. Matter {\bf
  20},  083203  (2008).

\bibitem{keldysh_1965_negf}
L. Keldysh, Sov. Phys. JETP {\bf 20},  1018  (1965).

\bibitem{Brandbyge02}
M. Brandbyge, J.~L. Mozos, P. Ordejon, and J.~T.~K. Stokbro, Phys. Rev. B {\bf
  65},  165401  (2002).

\bibitem{Toher05}
C. Toher, A. Filippetti, S. Sanvito, and K. Burke, Phys. Rev. Lett. {\bf 95},
  146402  (2005).

\bibitem{Choi07}
S.~G.~L. Hyoung Joon~Choi, Marvin L.~Cohen, Phys. Rev. B {\bf 76},  155420
  (2007).

\bibitem{beste_2008_basis_set_effect_transport}
A. Beste, V. Meunier, and R.~J. Harrison, J. Chem. Phys. {\bf 128},  154713
  (2008).

\bibitem{Calzolari04}
A. Calzolari, N. Marzari, I. Souza, and M.~B. Nardelli, Phys. Rev. B {\bf 69},
  035108  (2004).

\bibitem{Strange08}
M. Strange, I.~S. Kristensen, K.~S. Thygesen, and K.~W. Jacobsen, J. Chem.
  Phys. {\bf 128},  114714  (2008).

\bibitem{Delaney04}
P. Delaney and J.~C. Greer, Phys. Rev. Lett. {\bf 93},  036805  (2004).

\bibitem{Ferretti05}
A. Ferretti, A. Calzolari, R.~D. Felice, F. Manghi, M.~J. Caldas, M.~B.
  Nardelli, and E. Molinari, Phys. Rev. Lett. {\bf 94},  116802  (2005).

\bibitem{Cehovin08}
A. Cehovin, H. Mera, J.~H. Jensen, K. Stokbro, and T.~B. Pedersen, Phys. Rev. B
  {\bf 77},  195432  (2008).

\bibitem{Thygesen08}
K. Thygesen, Phys. Rev. Lett. {\bf 100},  166804  (2008).

\bibitem{Myohanen08}
P. My\"{o}h\"{a}nen, A. Stan, G. Stefanucci, and R. van Leeuwen, Europhysics
  Lett. {\bf 84},  67001  (2008).

\bibitem{Verdozzi06}
C.-O.~A. Claudio~Verdozzi, Gianluca~Stefanucci, Phys. Rev. Lett. {\bf 97},
  046603  (2006).

\bibitem{Frederiksen07}
T. Frederiksen, M. Paulsson, M. Brandbyge, and A.-P. Jauho, Phys. Rev. B {\bf
  75},  205413  (2007).

\bibitem{Galperin08}
M. Galperin, A. Nitzan, and M.~A. Ratner, Phys. Rev. B {\bf 78},  125320
  (2008).

\bibitem{bokes_2004_conductance}
P. Bokes and R.~W. Godby, Phys. Rev. B {\bf 69},  245420  (2004).

\bibitem{bokes_2007_four_point_conductance}
P. Bokes, J. Jung, and R. Godby, Phys. Rev. B {\bf 76},  125433  (2007).

\bibitem{Sai05}
N. Sai, M. Zwolak, G. Vignale, and M.~D. Ventra, Phys. Rev. Lett. {\bf 94},
  186810  (2005).

\bibitem{Jung07}
J. Jung, P. Bokes, and R.~W. Godby, Phys. Rev. Lett. {\bf 98},  259701  (2007).

\bibitem{Onida02}
G. Onida, L. Reining, and A. Rubio, Rev. Mod. Phys. {\bf 74},  601  (2002).

\bibitem{bruus_2004_many_body_book}
H. Bruus and K. Flensberg, {\em Many-body Quantum Theory in Condensed Matter
  Physics: An Introduction} (Oxford University Press, Oxford, 2004).

\bibitem{hohenberg_1964_DFT}
P. Hohenberg and W. Kohn, Phys. Rev. {\bf 136},  864  (1964).

\bibitem{kohn_1965_DFT_LDA}
W. Kohn and L.~J. Sham, Phys. Rev. {\bf 140},  A 1133   (1965).

\bibitem{ABINIT_short}
X. Gonze, J.-M. Beuken, R. Caracas, F. Detraux, M. Fuchs, G.-M. Rignanese, L.
  Sindic, M. Verstraete, G. Zerah, F. Jollet, M. Torrent, A. Roy, M. Mikami, P.
  Ghosez, J.-Y. Raty, and D.~C. Allan, Comp. Mat. Sci. {\bf 25},  478  (2002).

\bibitem{boeck_sfhingx}
S. Boeck, C. Freysoldt, A. Alsharif, A. Dick, L. Ismer, A. Qteish, and J.
  Neugebauer, http://www.sfhingx.de  .

\bibitem{troullier_1991_nc_psp}
N. Troullier and J.~L. Martins, Phys. Rev. B {\bf 43},  1993  (1991).

\bibitem{louie_1982_NLCC}
S.~G. Louie, S. Froyen, and M.~L. Cohen, Phys. Rev. B {\bf 26},  1738  (1982).

\bibitem{rojas_1995_GWST_prl}
H.~N. Rojas, R.~W. Godby, and R.~J. Needs, Phys. Rev. Lett. {\bf 74},  1827
  (1995).

\bibitem{maeland_1964_au_acell}
A. Maeland and T. Flanagan, Can. J. Phys. {\bf 42},  2364  (1964).

\bibitem{sanchez-portal_1999_stiff_monatomic}
D. Sanchez-Portal, E. Artacho, J. Junquera, P. Ordejon, A. Garcia, and J.
  Soler, Phys. Rev. Lett. {\bf 83},  3884  (1999).

\bibitem{hartwigsen_1998_psp_hgh}
C. Hartwigsen, S. Goedecker, and J. Hutter, Phys. Rev. B {\bf 58},  3641
  (1998).

\bibitem{marini_2001_Cu_GW}
A. Marini, G. Onida, and R. Del~Sole, Phys. Rev. Lett. {\bf 88},  016403
  (2001).

\bibitem{shishkin_2006_GW_PAW}
M. Shishkin and G. Kresse, Phys. Rev. B {\bf 74},  035101  (2006).

\bibitem{blochl_1994_PAW}
P.~E. Bl\"ochl, Phys. Rev. B {\bf 50},  17953  (1994).

\bibitem{Jelinek08}
P. Jel\'{i}nek, R. P\'{e}rez, J. Ortega, and F. Flores, Phys. Rev. B {\bf 77},
  115447  (2008).

\end{thebibliography}
\bibliographystyle{prsty}

\end{document}